\documentclass[%
reprint,
superscriptaddress,
amsmath,amssymb,
aps,
prb,
floatfix,
]{revtex4-2}

\usepackage{graphicx}
\usepackage{dcolumn}
\usepackage{bm}
\usepackage{hyperref}
\usepackage[mathlines]{lineno}
\usepackage{amsmath,bm}
\usepackage{times}
\usepackage{fouriernc}
\usepackage{xcolor}

\begin{document}

\preprint{APS/123-QED}

\title{Frustrated charge order and cooperative distortions in ScV$_6$Sn$_6$}

\author{Ganesh Pokharel}
\email{ganeshpokharel@ucsb.edu}
\affiliation{Materials Department, University of California Santa Barbara, California 93106, USA}

\author{Brenden R. Ortiz}
\affiliation{Materials Department, University of California Santa Barbara, California 93106, USA}

\author{Linus Kautzsch}
\affiliation{Materials Department, University of California Santa Barbara, California 93106, USA}

\author{Steven J. Gomez Alvarado}
\affiliation{Materials Department, University of California Santa Barbara, California 93106, USA}

\author{Krishnanand Mallayya}
\affiliation{Department of Physics, Cornell University, Ithaca, NY, 14853, United States}

\author{Guang Wu}
\affiliation{Department of Chemistry and Biochemistry, University of California Santa Barbara,  California 93106, USA}

 \author{Eun-Ah Kim}
\affiliation{Department of Physics, Cornell University, Ithaca, NY, 14853, United States}
\affiliation{Department of Physics, Ewha Womans University, Seoul, South Korea}

\author{Jacob P. C. Ruff}
\affiliation{CHESS, Cornell University, Ithaca, New York 14853, USA}

\author{Suchismita Sarker}
\affiliation{CHESS, Cornell University, Ithaca, New York 14853, USA}

\author{Stephen D. Wilson}
\email{stephendwilson@ucsb.edu}
\affiliation{Materials Department, University of California Santa Barbara, California 93106, USA}

\date{\today}

\begin{abstract}

Here we study the stability of charge order in the kagome metal ScV$_6$Sn$_6$.  Synchrotron x-ray diffraction measurements reveal high-temperature, short-range charge correlations at the wave vectors along \textbf{q}=($\frac{1}{3}$, $\frac{1}{3}$, $\frac{1}{2}$) whose inter-layer correlation lengths diverge upon cooling.  At the charge order transition, this divergence is interrupted and long-range order freezes in along \textbf{q}=($\frac{1}{3}$, $\frac{1}{3}$, $\frac{1}{3}$), as previously reported, while disorder enables the charge correlations to persist at the \textbf{q}=($\frac{1}{3}$, $\frac{1}{3}$, $\frac{1}{2}$) wave vector down to the lowest temperatures measured.  Both short-range and long-range charge correlations seemingly arise from the same instability and both are rapidly quenched upon the introduction of larger Y ions onto the Sc sites.  Our results validate the theoretical prediction of the primary lattice instability at \textbf{q}=($\frac{1}{3}$, $\frac{1}{3}$, $\frac{1}{2}$), and we present a heuristic picture for viewing the frustration of charge order in this compound.  
  
\end{abstract}

\pacs{Valid PACS appear here}
\maketitle

\section{Introduction}

Transition-metal-based kagome metals are of broad interest due to their potential for hosting a wide variety of electron correlation-driven instabilities and topologically nontrivial electronic states \cite{kang2020dirac, Yin2018, Yangeabb6003, Noam_2020, Pokharel_2021_2, Liu_2019, PhysRevLett.125.247002, Pokharel_2021, Tsai2020}. The three-sublattice geometry of corner-sharing triangles stabilizes hopping interference effects combined with band features endemic to hexagonal lattices \cite{PhysRevB.87.115135, kiesel2013unconventional}. Unconventional electronic states are theorized to arise in this setting, new, tunable classes of kagome metals are being actively sought to potentially realize these predictions.

One direction in this search is to explore, new nonmagnetic kagome variants, partially motivated by the discovery of superconductivity and unconventional charge density wave (CDW) in the $A$V$_3$Sb$_5$ ($A$=K, Rb, Cs) class of compounds \cite{ortiz2019new, PhysRevLett.125.247002}. Van Hove singularities (VHS) close to the Fermi level and derived from the saddle points arising from the V-based kagome network are thought to be key features driving their unusual properties. Recent studies have identified a class of $R$V$_6$Sn$_6$ ($R$= rare earth) that possess a similar band structure with VHS close to $E_F$ \cite{Pokharel_2021, Pokharel_2021_2}. While magnetic ions can be introduced onto the $R$-sites of these compounds to engineer a variety magnetic states \cite{Pokharel_3, PhysRevMaterials.6.083401, zhang2022electronic, rosenberg2022uniaxial}, the nonmagnetic variants, thus far, fail to stabilize a superconducting transition.  However, the variant with a smallest $R$-site cation radius, ScV$_6$Sn$_6$, was recently shown to manifest signatures of charge order \cite{Hasitha_2022}.  

The origin of the charge order instability in ScV$_6$Sn$_6$ has been the subject of several recent theoretical and experimental studies \cite{PhysRevB.107.165119, Binghai_2023, kang2023emergence, zhang2022destabilization}.  Charge order was first identified as a structural distortion below $T_{CO}=92$ K \cite{Hasitha_2022} where the lattice distorts along the wave vector \textbf{q}=($\frac{1}{3}$, $\frac{1}{3}$, $\frac{1}{3}$) and the symmetry is lowered from $P6/mmm$ to $R32$.  In contrast to the in-plane kagome breathing mode that dominates the charge density wave state in the $A$V$_3$Sb$_5$ compounds \cite{kautzsch2023structural, christensen2021theory}, in the primary distortion in ScV$_6$Sn$_6$ is an out-of-plane modulation of the Sn and Sc sites.  Multiple recent studies have identified that this instability arises from strong electron-phonon coupling \cite{tuniz2023dynamics, hu2023phonon}; however recent DFT models have identified that the most energetically favorable distortion mode has a different inter-layer periodicity relative to the distortion experimentally observed \cite{Binghai_2023}.  Curiously, a number of nearly degenerate distortion wave vectors were identified, suggesting that competition between these states may be observed in the dynamics or short-range order (diffuse scatter) of this material.

Here we explore this concept by parameterizing the evolution and stability of the charge ordered state of ScV$_6$Sn$_6$ via synchrotron x-ray diffraction measurements and isoelectronic doping. We observe two, coexisting, charge order wave vectors with propagation \textbf{q$_{\frac{1}{3}}$}=($\frac{1}{3}, \frac{1}{3}, \frac{1}{3}$) and \textbf{q$_{\frac{1}{2}}$}=($\frac{1}{3}, \frac{1}{3}, \frac{1}{2}$). More robust, yet short-ranged charge correlations are observed at all temperatures (15 K < $T$ < 300 K) at \textbf{q$_{\frac{1}{2}}$}, corresponding to the energetically favored mode predicted by Density Function Theory (DFT). These short-range correlations are frustrated from forming long-range order via the kagome motif and strengthen upon cooling. The inter-layer correlation length $\xi_c$ diverges as the onset of charge order ($T_{\text{co}}$) is approached, and once $\xi_c$ reaches a value of approximately three $c$-axis lattice constants, the first-order onset of long-range \textbf{q$_{\frac{1}{3}}$} charge correlations occurs. Continued cooling reveals the coexistence of both \textbf{q$_{\frac{1}{3}}$} and \textbf{q$_{\frac{1}{2}}$} charge correlations in the ground state. Isoelectronic doping of larger Y ions into (Sc$_{1-x}$Y$_x$)V$_6$Sn$_6$ rapidly suppresses the charge ordered state, and our data demonstrate that long-range \textbf{q$_{\frac{1}{3}}$} correlations vanish by $x\approx 0.03$ and short-range \textbf{q$_{\frac{1}{2}}$} correlations vanish by a critical value $x\approx 0.08$.  In aggregate, our data demonstrate an unusual cooperative phenomenology involved in the formation of charge order in ScV$_6$Sn$_6$  where \textbf{q$_{\frac{1}{3}}$} and \textbf{q$_{\frac{1}{2}}$} correlations arise from a common lattice instability.


\section{Experimental Details}
\subsection{Crystal synthesis}
Single crystals of Sc$_{1-x}$Y$_x$V$_6$Sn$_6$ (x=0, 0.03, 0.08) were grown via a conventional flux-based growth technique. Sc (pieces, 99.9\%), Y (powder, 99.5\%), V (pieces, 99.7\%), and Sn (shot, 99.99\%)  were loaded inside a Canfield crucible with various molar ratios depending upon the Sc doping and then sealed in a quartz tube filled with 1/3 atmosphere of Ar. For Sc$_{1-x}$Y$_x$V$_6$Sn$_6$, more Y tends to incorporate than the nominal doping concentration, reflective of the fact that YV$_6$Sn$_6$ is more structurally stable relative to ScV$_6$Sn$_6$. As a result, the starting elemental ratios of Sc, Y, V, and Sn were selected to be 1:0:6:50, 0.99:0.01:6:50, 0.93:0.03:6:50 to realize doping levels of x = 0, 0.03, 0.08 respectively. Sealed quartz tubes loaded with the element mixtures were heated at 1150$^\circ$C for 12 hours. Then, the furnace was slowly cooled down at a rate of 2$^\circ$C/h to 780 $^\circ$C, and faceted, plate-like single crystals were separated from the molten Sn flux via centrifuging at 780 $^\circ$C. Crystals were typically 1-2 mm in lateral length and near 500 $\mu m$ in thickness. The surfaces of some crystals were occasionally contaminated with excess flux Sn and were subsequently cleaned with dilute HCl to remove the surface flux.

\begin{figure*} 
\centering
\includegraphics[scale=0.5]{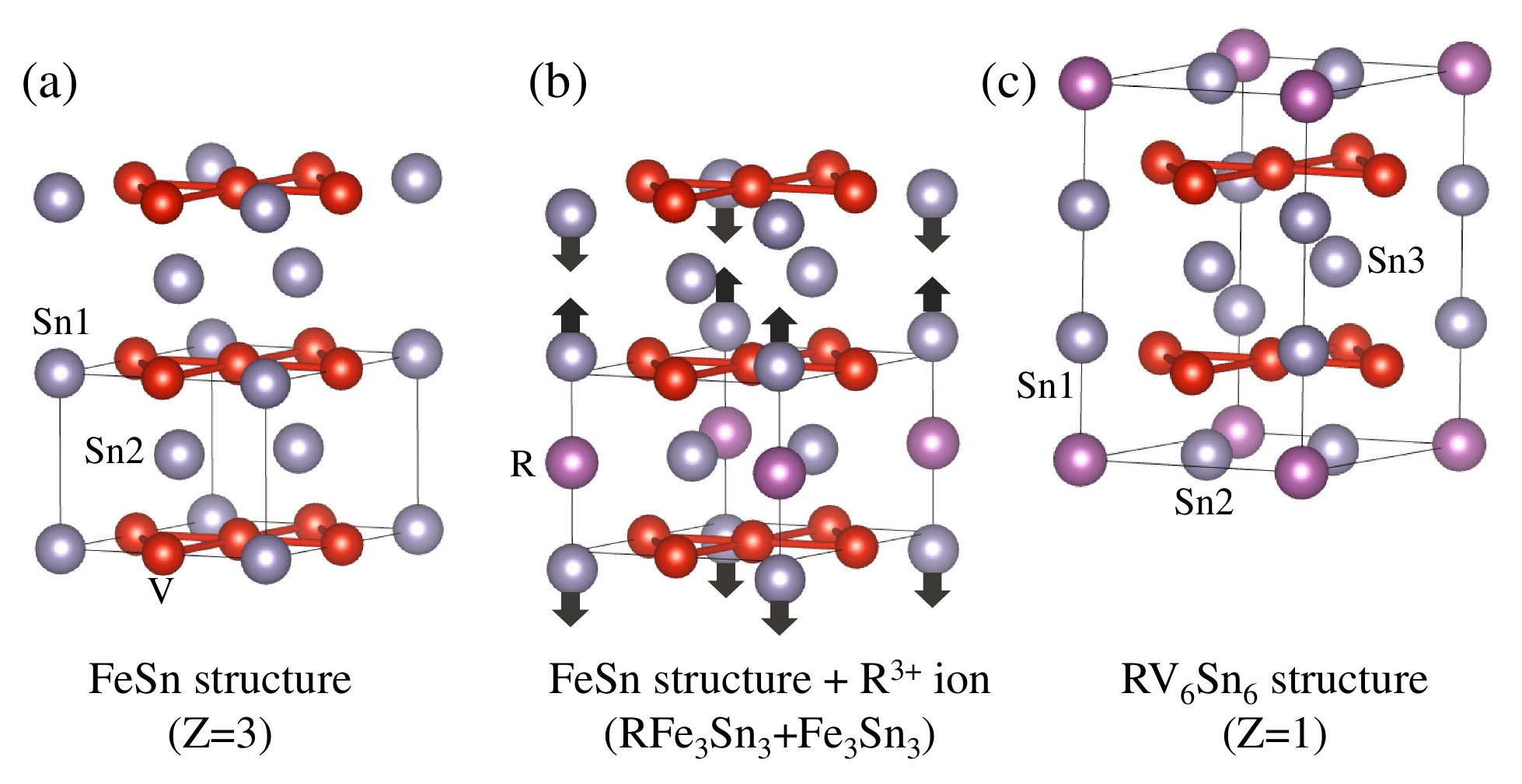}
\caption{Structural motif of $R$V$_6$Sn$_6$ ($R$= rare earth) compounds and its relation to the FeSn structure type.  (a) FeSn structure-type of hypothetical VSn showing a V kagome net (red spheres) with Sn1 sites (grey spheres) located in the kagome plane within the hexagon voids of the kagome lattice.  In between kagome planes, Sn2 (grey spheres) ions form a honeycomb network.  (b) The FeSn structure type modified with $R^{3+}$ ions added into the voids of the honeycomb lattice of the interstitial Sn2 network.  The larger size of the $R^{3+}$ ion pushes the $c$-axis adjacent Sn1 sites out of the kagome plane.  This vertical displacement sterically biases the insertion of $R$ ions into every other plane of honeycomb Sn2, resulting in a distinct interstitial Sn3 stannene layer. (c) The resulting $R$V$_6$Sn$_6$ unit cell recentered on the new $R^{3+}$ ions revealing the resulting bilayer structure.
}
\label{struct_evo}
\end{figure*}

\begin{figure*} 
	\centering
	\includegraphics[scale=0.6]{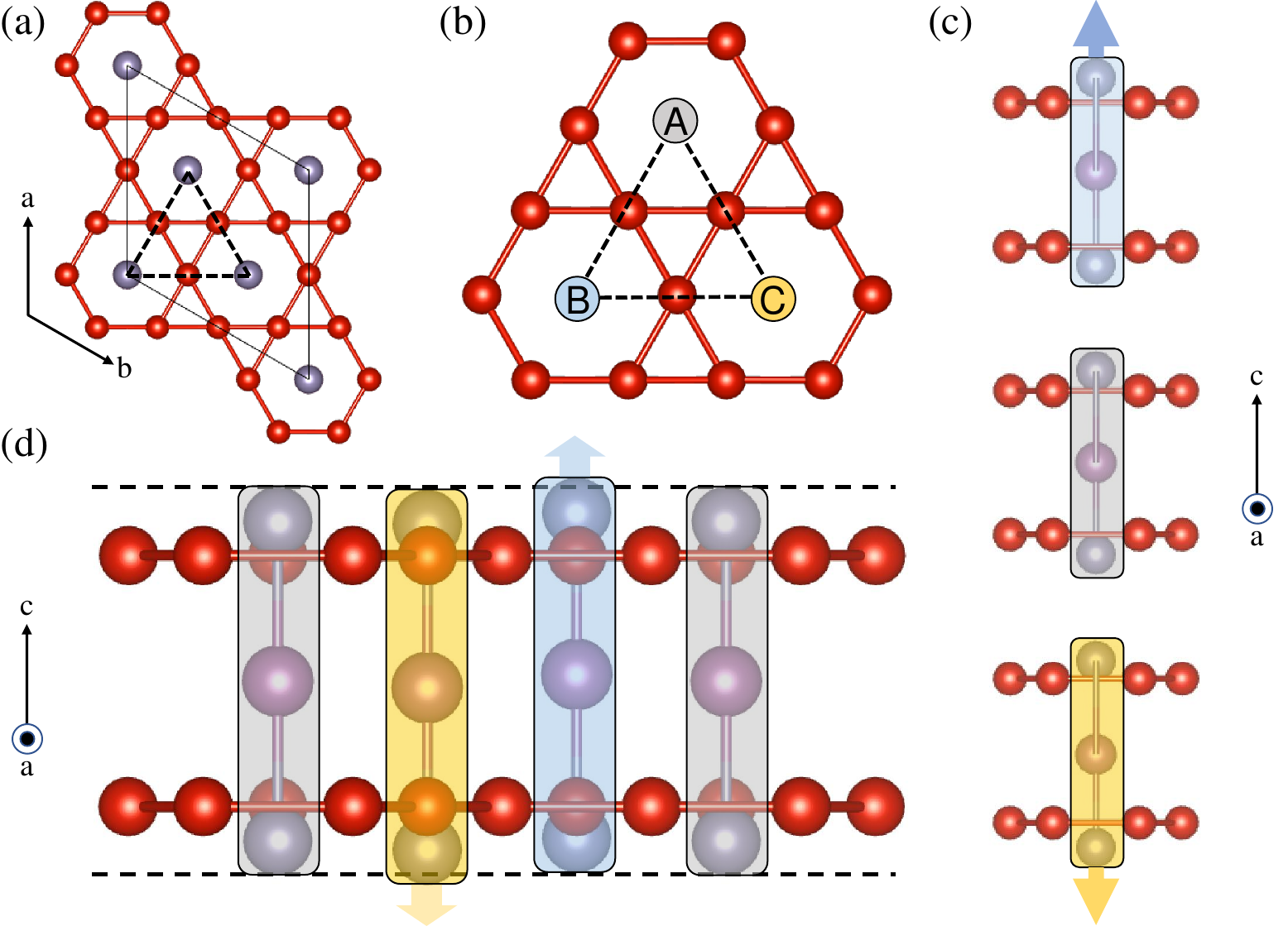}
	\caption{Illustration of the frustrated lattice distortion in ScV$_6$Sn$_6$. (a) shows the $ab$-plane with only V and Sn1 sites populated.  The Sn1 sites sit directly above/below the V planes and centered over the voids of the hexagons in the kagome lattice.  The network of these hexagon centers forms a triangular lattice.  (b)  The triangular lattice hexagons threaded by the Sn1-Sc-Sn1 trimers.  The charge ordered state decorates this lattice with three types of trimer shifts A (neutral), B (up), and C (down) in order to minimize the energy cost of bringing the Sn1 sites closer to the V plane.  (c) shows the corresponding trimer shifts along the $c$-axis chain threading through a given hexagon (d) $bc$-plane showing the trimer shifts and relative displacements within a given bilayer pair.
	}
	\label{distortion}
\end{figure*}

\subsection{Scattering measurements}
Laboratory single-crystal x-ray diffraction data were collected at room temperature on a Kappa Apex II single-crystal diffractometer with a Mo source and a fast charge-coupled device (CCD) detector. Data were refined to obtain the structural solution using SHELX software package \cite{Shelx_2015}. Temperature-dependent synchrotron x-ray diffraction experiments were performed on the ID4B (QM2) beamline at the CHESS light source. To avoid the absorption edges of constituent elements, an incident x-ray of wavelength  $\lambda =$ 0.676 $\AA$ ($E= 26$ KeV) was chosen using a double-bounce diamond monochromator.  A stream of cold helium gas flowing across the crystal sample was used to control the temperature. 

An area detector array collected the scattering intensities in transmission geometry, and data were collected while the samples were rotated with three tilted 360$^\circ$ rotations, sliced into 0.1$^\circ$ frames. A Pilatus 6M detector was used to collect the data.  Early-stage data was visualized and then analyzed using the NeXpy software package. Final structural analysis of the diffraction data was performed using the APEX3 software package \cite{Apex_53487} where the structures at various temperatures were refined using the integrated SHELX and APEX3 software package. An unsupervised machine learning tool using Gaussian Mixture Models called X-ray Temperature Clustering (X-TEC)~\cite{Venderley2022Proc.Natl.Acad.Sci.} was used to identify classes of peaks in reciprocal space with distinct temperature dependencies.  The viewing axes for the ($H$, $K$) scattering planes shown in this paper are skewed by 60$^{\circ}$ to reflect the peaks corresponding to the hexagonal lattice system and this is illustrated by the white dotted lines in the top left panel of Figure 3.  

\subsection{Magnetization and heat capacity measurements}
Magnetization measurements were performed as a function of temperature and magnetic field using a Quantum Design Magnetic Properties Measurement System (MPMS3). To remove any surface contamination prior to the measurements, the plate-like single crystals were polished gently on the top and bottom surfaces. Crystals of mass $\sim$ 1 mg were attached to quartz paddles using GE-Varnish and then loaded in the MPMS. Magnetization data were collected in the temperature range of 2 K to 300 K with a magnetic field of $\mu_0 H$=1 T applied parallel to the crystal surface. Heat capacity measurements were performed on Sc$_{1-x}$Y$_x$V$_6$Sn$_6$ single crystals of a similar mass ($\approx$ 1 mg) were carried out using a Dynacool Physical Properties Measurement System (PPMS). The samples were mounted to the addenda using N-grease, and heat capacity measurements were carried out in same temperature range of 2 K to 300 K under zero magnetic field. 

\section{Results}

\subsection{Heuristic picture of the lattice distortion in ScV$_6$Sn$_6$}
A lattice distortion signifying charge order was reported in ScV$_6$Sn$_6$ \cite{Hasitha_2022}, and we performed sychrotron x-ray diffraction measurements parameterizing the average structure at two temperatures, $T=30$ K (below $T_{\text{co}}$) and $T=$300 K (above $T_{\text{co}}$).  The results of these average structure measurements are consistent with those previously reported and will be used to frame the discussion in this section. 

ScV$_6$Sn$_6$ contains a much smaller $R$-site ion relative to other $R$V$_6$Sn$_6$ variants such as YV$_6$Sn$_6$, yet it still maintains a qualitatively similar electronic band structure \cite{Binghai_2023}.  Steric effects and modified bonding associated with the size of Sc are therefore strongly suggested to play a role in the formation of the charge ordered state.  The lattice distortion observed below $T_{\text{co}}$ has a reported wave vector \textbf{q$_{\frac{1}{3}}$}=($\frac{1}{3}$, $\frac{1}{3}$, $\frac{1}{3}$), implying a unit cell that is enlarged by $\sqrt(3)\times\sqrt(3)\times3$.  This differs from the predicted DFT cell with \textbf{q$_{\frac{1}{2}}$}=($\frac{1}{3}$, $\frac{1}{3}$, $\frac{1}{2}$) and size $\sqrt(3)\times\sqrt(3)\times2$.  Below we discuss the elements of the $R$V$_6$Sn$_6$ structure-type and propose a heuristic picture of frustrated charge order in this compound that stabilizes the $\sqrt(3)\times\sqrt(3)\times3$ state. This picture is built around the crucial role of the Sn1-$R$-Sn1 unit and its frustrated interaction with the kagome plane through $T_{\text{co}}$ in ScV$_6$Sn$_6$.


Figure 1 shows the key elements of the $R$V$_6$Sn$_6$ structure.  Figure 1 (a) starts from a theoretical structure built from a base FeSn-type lattice and populated with V on the Fe sites.  This lattice contains Sn on two distinct sites, the first is Sn1 in the kagome plane and centered inside the hexagonal-voids of the kagome network.  The second, Sn2 site, places Sn within a honeycomb network between the kagome planes and arranged such that the Sn2 sites sit above and below the triangles of the kagome network.  This structure can be evolved by adding $R^{3+}$ ions within the hexagonal voids of the interstitial Sn2 honeycomb network as shown in Figure 1 (b).  The larger $R$-site ion pushes the Sn1 sites away from the $R$-site center and out of the kagome plane.  This steric effect, combined with the $R$-Sn1 bonding is a key element of the modified structure. $R$-sites occupy every other stannene layer and create a unique Sn3 stannene plane.  The resulting, enlarged $R$V$_6$Sn$_6$ unit cell with a new $R$-site center is shown in Figure 1 (c).

If one considers the structural modifications upon going from YV$_6$Sn$_6$ to ScV$_6$Sn$_6$, one difference is a relative contraction of the Sn1-$R$-Sn1 bond lengths which shorten from Y-Sn1=0.3335$c$ to Sc-Sn1=0.3238$c$ at 300 K, where $c$ is the $c$-axis lattice parameter. This also results in a contraction of the vertical displacement of the Sn1 site ions relative to the kagome plane. At 300 K, Sn1 sites in YV$_6$Sn$_6$ sites show a vertical displacement of (Sn1-V$_{\text{plane}}$)=0.0854$c$ relative to the middle of the V-hexagons, while in ScV$_6$Sn$_6$ they are substantially closer at (Sn1-V$_{\text{plane}}$)=0.077$c$. 

When viewing the longer chain of $R$-Sn1-Sn1-$R$ along $c$, the $R$-Sn1 contraction effectively breaks up the equally spaced chain of 0.334$c$-0.333$c$-0.334$c$ in YV$_6$Sn$_6$ into a pseudodimerized chain of 0.324$c$-0.352$c$-0.324$c$ in ScV$_6$Sn$_6$. The bond distances along the units of these chains change from having only a $0.15\%$ modulation to a $4\%$ modulation about their mean values.  These segmented and weakly bound Sn1-Sc-Sn1 ``trimers" within the disrupted chain are now free to respond to steric effects associated with the proximity of Sn1 atoms to the kagome plane.  In other words, the stronger bonding within the $R$-Sn1-Sn1-$R$ chains is inside the Sn1-$R$-Sn1 trimer units, and the metal-metal bonding between the Sn1-Sn1 sites is weakened in ScV$_6$Sn$_6$.  The result is a chain that segments via a Peierls-like distortion into two trimerized Sn1-Sc-Sn1 units that run through the hexagonal voids in the kagome plane.

By viewing the lattice in terms of these trimers, one can empirically analyze what structural impediment is relieved upon cooling below $T_{\text{co}}$.  Comparing the 300 K data to 30 K data, the average Sc-Sn1 bond length comprising the trimers is more or less unaffected (contracting $<0.5\%$).  The average (Sn1-V$_{\text{plane}}$) distance is, however, strongly impacted, and it contracts by $6.3\%$.  This indicates that there is an energy penalty for the Sn1 sites approaching the kagome plane that the \textbf{q$_{\frac{1}{3}}$} charge order relieves.  Below $T_{\text{co}}$, the V atoms displace by $\approx0.13$ $\AA$ along $c$ and only by 0.02 $\AA$ to 0.06 $\AA$ within the plane, consistent with the Sn1-V interaction playing an important role in the low temperature distortion.

A picture of energetically favored trimerization pulling Sn1 units inward (toward Sc) competing with energetically disfavored Sn1 atoms approaching both V planes within a bilayer (see Figure 1 (c)) necessarily supports frustration within the structure.  Figure 2 (a) illustrates the $ab$-plane of ScV$_6$Sn$_6$ with only the V and Sn1 atoms shown.  This shows the network of hexagons that the Sn1 atoms approach in the kagome plane, which in turn form a triangular lattice.  As the Sc-Sn1-Sn1-Sc chains running through these hexagon centers are broken up due to the diminished Sn1-Sn1 bonds, the lattice has an Ising-like degree of freedom to address the frustration via displacement of the Sn1-Sc-Sn1 trimers along $c$.  One way of attempting to do this is via neighboring trimers  shifting with opposite senses, which is the DFT predicted \textbf{q$_{\frac{1}{2}}$} structure.  This solution naively has an extensive degeneracy when mapped to an antiferromagnetic triangular lattice Ising model, and the distortion should result in significant fluctuations and residual entropy.  Given that the segmentation of the Sc-Sn1-Sn1-Sc chains has internal degrees of freedom, this residual entropy can be relieved via an alternative, longer wavelength modulation along the chain.

\begin{figure*} 
	\centering
	\includegraphics[scale=0.55]{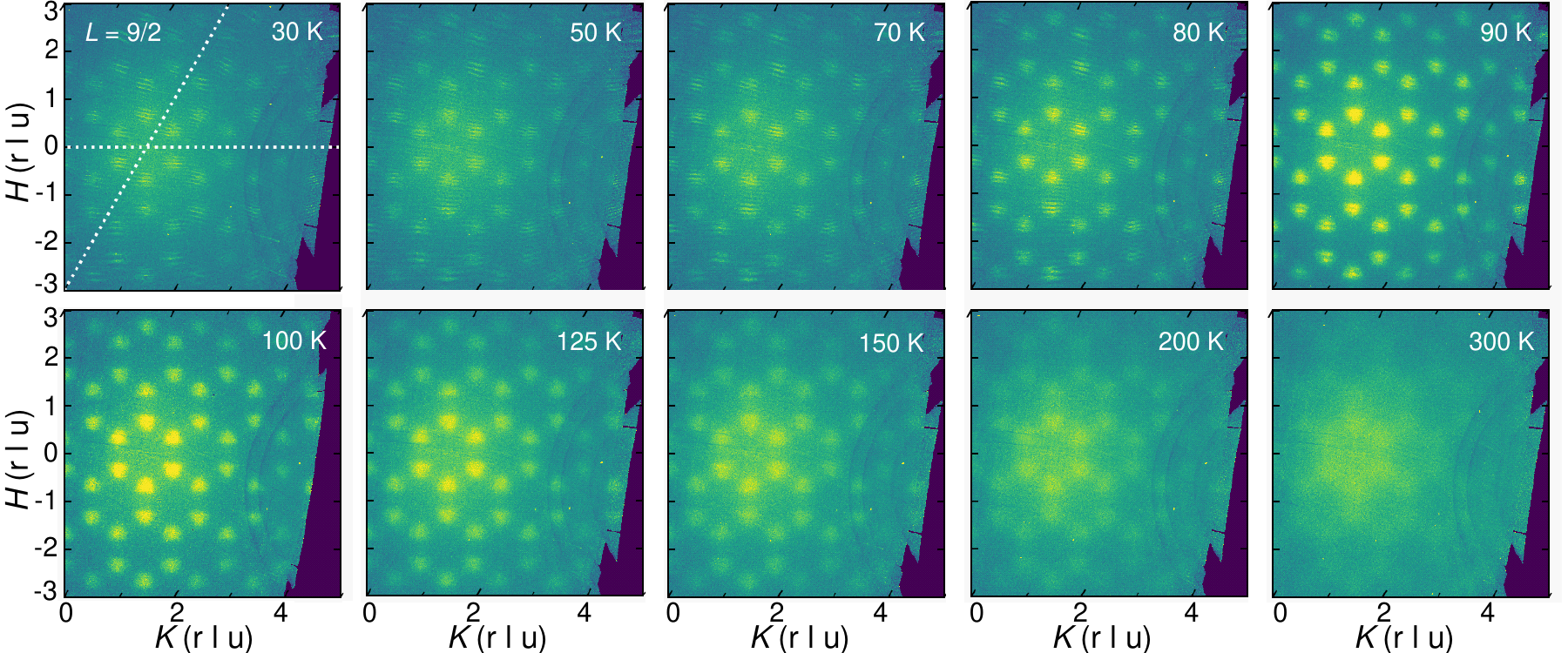}
	\caption{Synchrotron x-ray diffraction data showing the temperature dependence of charge correlations in the $L=\frac{9}{2}$ plane. Axis units are plotted in reciprocal lattice units (r l u), and the white dotted lines in the top left panel illustrate the ($H$, 0, $\frac{9}{2}$) and (0, $K$, $\frac{9}{2}$) paths. Data collected from 300 K through 30 K are plotted revealing the evolution of short-range charge correlations at ($\frac{1}{3}$, $\frac{1}{3}$, $\frac{1}{2}$)-type positions upon cooling through $T_{\text{co}}\approx84$ K in this sample.
	}
	\label{}
\end{figure*}

The simplest way to envision a chain modulation that lifts the frustration fully is to consider the experimentally realized distortion at \textbf{q$_{\frac{1}{3}}$}.  As shown in Figures 2 (b) and (c), three distortion types labeled A, B, and C can solve the frustration by having one trimer shift up, one shift down, and the other not shift at all (neutral). This would map the frustration into a 3-state Potts model with antiferromagnetic interactions, and, for dominant nearest-neighbor interactions, this model has a first order transition into long-range order \cite{Enting1982}. The ABC ordered state shown in Figure 2 (b) is the ground state of such a model in the limit where next-nearest neighbor interactions vanish \cite{murtazaev2016three}.  Figure 2 (d) shows the resulting pattern of trimer displacements about a given kagome bilayer plane.  This long-range charge ordered state is the one realized in ScV$_6$Sn$_6$.  Next, we discuss x-ray diffraction results that highlight presence of substantial fluctuation effects at \textbf{q$_{\frac{1}{2}}$} consistent with this picture and that both the fluctuations and long-range order at \textbf{q$_{\frac{1}{3}}$} arise from a common instability.

\subsection{Thermal evolution of frustrated q$_{\frac{1}{2}}$ correlations}

Looking first at the wave vector predicted to be the most stable via DFT, Figure 3 shows the thermal evolution of charge scattering at \textbf{q$_{\frac{1}{2}}$}.  Starting at room temperature, weak, short-ranged charge correlations are evident at 300 K that strengthen upon cooling.  These correlations remain short-range and centered at the frustrated \textbf{q$_{\frac{1}{2}}$} wave vector, and notably they persist both above and below $T_{\text{co}}$.  Analysis of the intensity of these short-range \textbf{q$_{\frac{1}{2}}$} correlations reveals that they reach a maximum near $T_{\text{co}}$ and then partially condense into the long-range ordered state.  The weak ripple-like modulations in the scattering intensities observed at low temperature likely arise from an artifact of unknown origin, and we believe they were introduced upon switching over to He-gas cooling below 100 K.  Repeat measurements at low-temperature on a second crystal did not reproduce these intensity modulations. 

Further illustrating the interplay of both correlation types are plots shown in Figure 4 within the charge ordered state ($T =$ 15 K) and within the high-temperature, disordered state ($T= 300$ K).  Deep within the ordered state, sharp superlattice peaks are evident in Figure 4 (a) in the $L=\frac{8}{3}$ plane, and a triangular network of diffuse scattering in the frustrated $L=\frac{9}{2}$ plane is present as shown in Figure 4 (c). Diffuse scatter from the short-range inter-layer correlation lengths associated with \textbf{q$_{\frac{1}{2}}$} correlations is also evident between the sharp \textbf{q}=($\frac{1}{3}$, $\frac{1}{3}$)-type satellites in the $L=\frac{8}{3}$ plane of Figure 4 (a). At temperatures far above $T_{\text{co}}$, Figure 4 (b) shows that correlations vanish at \textbf{q$_{\frac{1}{3}}$}-type positions yet weakly persist at \textbf{q$_{\frac{1}{2}}$} wave vectors.

\begin{figure} 
	\centering
	\includegraphics[width=1\columnwidth]{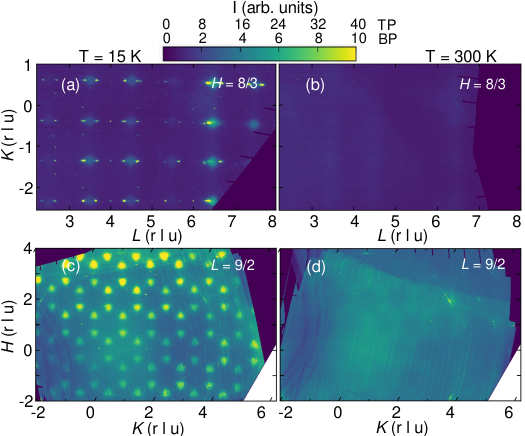}
	\caption{Synchrotron x-ray diffraction data showing a comparison of charge correlations in $H=\frac{8}{3}$ and $L=\frac{9}{2}$ scattering planes. Axis units are plotted in reciprocal lattice units (r l u). (a)  Select long-range \textbf{q$_{\frac{1}{3}}$} superlattice reflections are shown in the ($\frac{8}{3}$, $K$, $L$) plane at 15 K. (b) The same ($\frac{8}{3}$, $K$, $L$) scattering plane is shown at 300 K. (c) \textbf{q$_{\frac{1}{2}}$}-type short-range correlations are shown in the ($H$, $K$, $\frac{9}{2}$) plane at 15 K. (d) The same ($H$, $K$, $\frac{9}{2}$) scattering plane is shown at 300 K. The intensities of scattering data in the top panels (TP) for the $H=\frac{8}{3}$ plane and the bottom panels (BP) for the $L=\frac{9}{2}$ plane are different and they are represented by two different scales in the intensity bar.
	}
	\label{}
\end{figure}

\begin{figure}
	\includegraphics[width=0.985\linewidth]{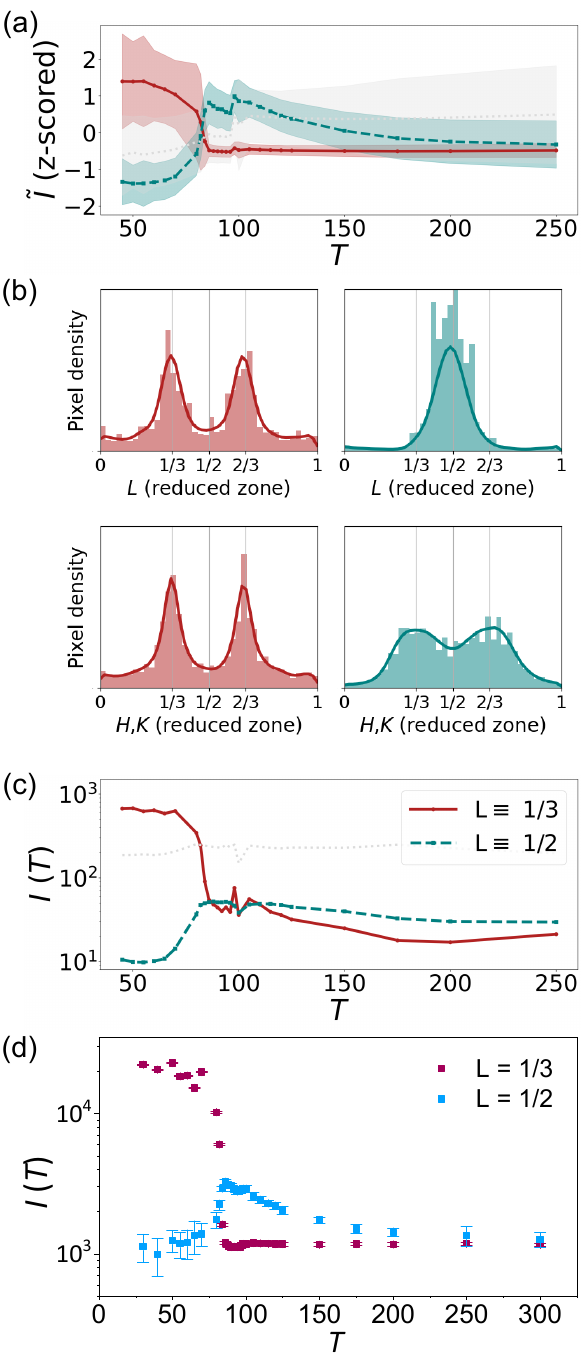}
	\caption{X-TEC analysis of ScV$_6$Sn$_6$ diffraction data. (a) X-TEC identifies two distinct temperature trajectories as red and teal clusters. The lines denote the mean, and shading denotes one standard deviation of the rescaled (z-scored) intensities in each cluster. The grey cluster captures the relatively temperature-independent pixels of the background. (b) The momentum distribution (in the reduced Brillouin zone) of the pixels in the red and teal clusters. (c) The average intensity of the red and teal cluster reveals the first-order transition of the $\mathbf{q}_{1/3}$ order at $T_{\text{co}}\approx$84 K and the evolution of $\mathbf{q}_{1/2}$ short-range order respectively. (d). Temperature-dependent integrated peak intensity, revealing an evolution of $\mathbf{q}_{1/3}$ and $\mathbf{q}_{1/2}$-type superlattice peaks above and below $T_{\text{co}}$. }
	\label{Fig_XTEC}
\end{figure}

The detailed temperature dependence of intensities at \textbf{q$_{\frac{1}{2}}$} and \textbf{q$_{\frac{1}{3}}$} positions are plotted in Figure 5. Analysis of the temperature series X-ray data for ScV$_6$Sn$_6$ is enabled by the unsupervised machine learning tool called X-ray Temperature Clustering (X-TEC)~\cite{Venderley2022Proc.Natl.Acad.Sci.}. X-TEC uses a Gaussian Mixture Model to identify the intensities in reciprocal space with distinct temperature dependencies. In order to distinguish the functional form of the intensity-temperature trajectory rather than their magnitudes, the intensity of each momentum $\vec{q}$ is re-scaled (z-scored) as $\tilde{I}_{\vec{q}}(T)=\left(I_{\vec{q}}(T)-\mu_{\vec{q}}\right)/\sigma_{\vec{q}}$ where $\mu_{\vec{q}}$ is the mean over temperature $T$, and $\sigma_{\vec{q}}$ is the standard deviation in $T$~\cite{Venderley2022Proc.Natl.Acad.Sci.}. By applying X-TEC on the entire data (37GB), we identify two distinct temperature trajectories color-coded as red and teal clusters, as shown in Figure~\ref{Fig_XTEC} (a). The distribution of these clusters in the reciprocal lattice (Figure~\ref{Fig_XTEC} (b)) shows that pixels of the red cluster are sharply concentrated at the ($\frac{1}{3}$, $\frac{1}{3}$, $\frac{1}{3}$) position corresponding to the $\mathbf{q}_{1/3}$ order, while the teal cluster pixels are broadly distributed around  ($\frac{1}{3}$, $\frac{1}{3}$, $\frac{1}{2}$), corresponding to the short-ranged $\mathbf{q}_{1/2}$ order. The average intensity of the red and teal cluster (Figure~\ref{Fig_XTEC} (c)) reveals the interplay of the two correlation types, as well as the onset of charge order at $\approx$84 K.  This combined with the development of the z-scored intensity $\tilde{I}$ shown in Figure 5 (a) supports the idea that \textbf{q$_{\frac{1}{2}}$} correlations develop toward an instability that is thwarted by the first-order onset of the $\mathbf{q}_{1/3}$ state. This trade-off in intensity is further illustrated in Fig. 5 (d) showing a conventional integration of the  ($\frac{1}{3}$, $\frac{4}{3}$, $\frac{13}{3}$) (L=1/3) and ($\frac{2}{3}$, $\frac{2}{3}$, $\frac{9}{2}$) (L=1/2) superlattice reflections as a function of temperature.

The interplay between these correlation types was further parameterized by resolution-convolved Lorenztian fits to the in-plane and out-of-plane peak profiles, and select examples are shown in Figure 6.  Consistent with the first-order nature of the charge order transition at \textbf{q$_{\frac{1}{3}}$}, Figure 6 (a) shows that peaks at this wave vector are sharp and resolution-limited (i.e. limited by the crystallinity of the material) immediately after appearing below $T_{\text{co}}$.  The poorer inter-layer crystallinity likely arises from extrinsic structural disorder in the sample that also slightly depresses $T_{\text{co}}$. Correlations at the frustrated wave vector \textbf{q$_{\frac{1}{2}}$}, however, evolve continuously upon cooling.  

Fit to peaks centered at \textbf{q$_{\frac{1}{2}}$} are shown in Figures 6 (a) and (b) with the resulting correlation lengths plotted in Figure 6 (c). The correlation length was determined by fitting peaks to Voigt line shapes with the Gaussian component determined by a nearby, primary Bragg reflection.  The resulting convolved Lorentzian full width at half maximum $\omega$ (in units of ~\AA$^{-1}$ using $Q=\frac{2\pi}{d}$) was then used to determine the correlation length $\xi=\frac{2}{\omega}$. 

At 300 K, inter-layer correlations are present with $\xi_c$ between 1 and 2 unit cells, while in-plane correlations are seemingly frustrated and absent.  Upon cooling, interlayer correlations begin to diverge and in-plane correlations begin to be resolved below 200 K.  Continued cooling drives enhanced fluctuations and a smooth increase in both $\xi_c$ and $\xi_{ab}$ until $\xi_c$ reaches 3 unit cells.  When $\xi_c=3c$, frustration is relieved via the long-range modulation of the Sn1-Sc-Sn1 trimers and the onset of \textbf{q$_{\frac{1}{3}}$} order.  Fluctuations at \textbf{q$_{\frac{1}{2}}$} are simultaneously reduced. Below $T_{\text{co}}$, $\xi_c$ is reduced to $\approx$ 2 unit cells, suggesting there is a fraction of the sample that does not undergo the transition into long-range order (likely due to disorder, as discussed later) or, alternatively, due to softening and hardening of the lattice dynamics about $T_{\text{co}}$.  This evolution is consistent with a frustrated lattice instability whose frustration is relieved when inter-layer coherence reaches the 3 $c$-axis lattice constants necessary to unlock an additional internal degree of freedom.  This internal degree of freedom is the additional trimer modulation along the $c$-axis chain discussed in the previous section.

\begin{figure} 
	\centering
	\includegraphics[scale=0.55]{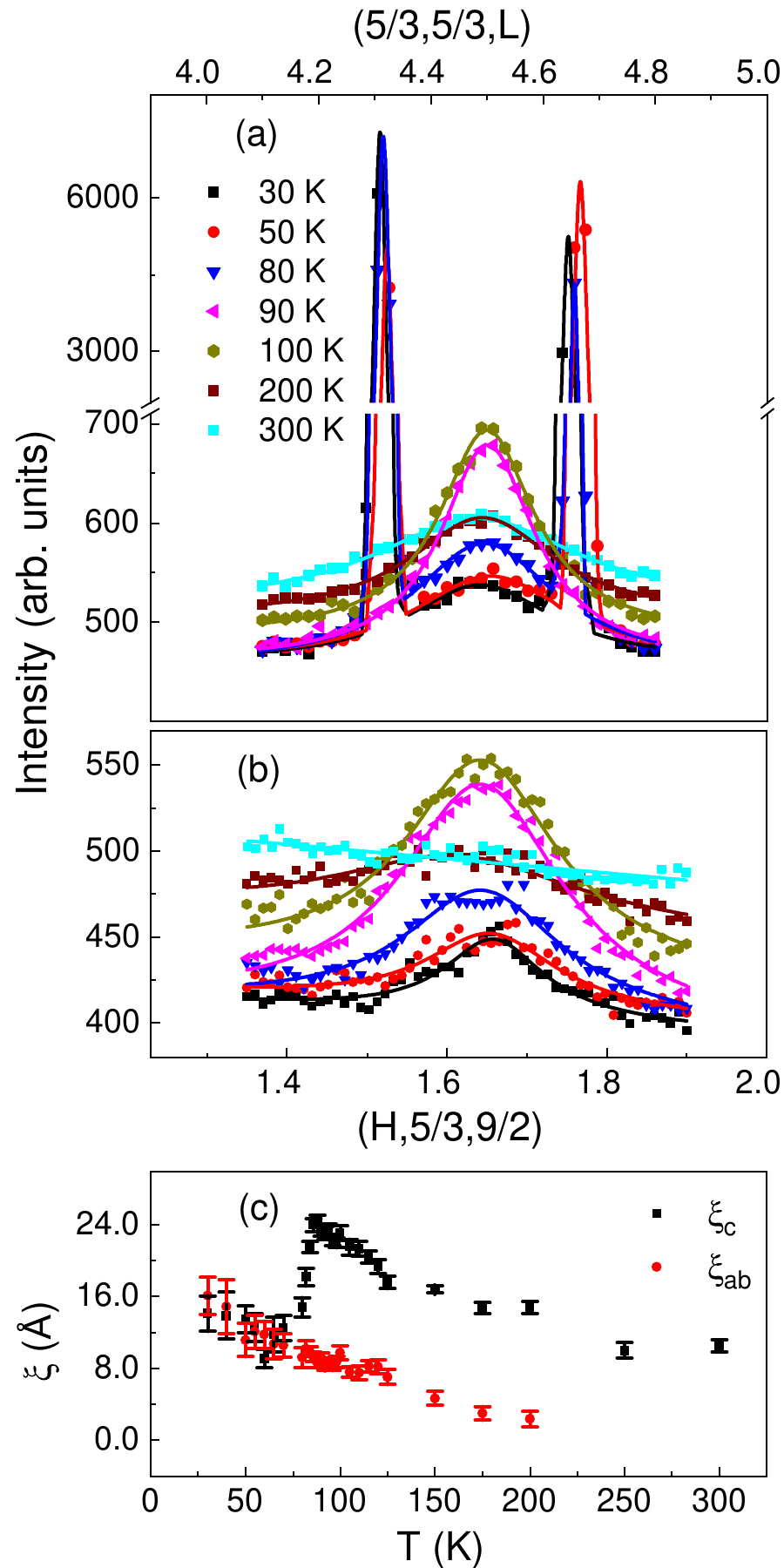}
	\caption{(a) $L$-scans used to parameterize $\xi_c$($T$) for \textbf{q$_{\frac{1}{2}}$} and \textbf{q$_{\frac{1}{3}}$} orders.  Data are fit through the ($\frac{5}{3}$, $\frac{5}{3}$, $\frac{13}{3}$) and ($\frac{5}{3}$, $\frac{5}{3}$, $\frac{9}{2}$) positions at various temperatures. (b) $H$-scans used to parameterize $\xi_{ab}$($T$) for \textbf{q$_{\frac{1}{2}}$} order. Data are fit through the ($\frac{5}{3}$, $\frac{5}{3}$, $\frac{9}{2}$) position at various temperatures. (c) Temperature dependence of $\xi_{ab}$ and $\xi_{c}$ for \textbf{q$_{\frac{1}{2}}$} charge correlations extracted via fits to peaks at the ($\frac{5}{3}$, $\frac{5}{3}$, $\frac{9}{2}$) position.  
	}
	\label{}
\end{figure}

\subsection{Response of frustrated charge order to perturbation of $c$-axis chains}

 ScV$_6$Sn$_6$ is the only known member of the family of $R$V$_6$Sn$_6$ materials that undergoes a charge order instability \cite{Pokharel_2021, Pokharel_2021_2, Pokharel_3, PhysRevMaterials.6.083401,zhang2022electronic}.  To probe the response of charge order to a changing $R$-site radius, the evolution of both long-range \textbf{q$_{\frac{1}{3}}$} and short-range \textbf{q$_{\frac{1}{2}}$} correlations was explored as a function of alloying Y onto the Sc site. Three Y-doped samples were grown with Sc$_{1-x}$Y$_x$V$_6$Sn$_6$ ($x=$ 0, 0.03, 0.08) and characterized initially via susceptibility and heat capacity measurements.

Results from bulk susceptiblity and heat capacity measurements of Sc$_{1-x}$Y$_x$V$_6$Sn$_6$ are shown in Figure 7.  Susceptibility data show that light substitution of Y suppresses $T_{\text{co}}$, and it reduces from 89 K in $x=0$ down to 85 K in the lightly $x=0.03$ sample.  The anomaly vanishes in a first-order like fashion in the $x=0.08$ concentration.  This thermal evolution is further illustrated in the inset of Figure 7 which shows the heat capacity, C$_p$($T$) of Sc$_{1-x}$Y$_x$V$_6$Sn$_6$. A sharp anomaly appears in the $x=0$ compound near 87 K, and the sharp transition weakens significantly and shifts downward with added Y disorder. Charge order vanishes by $x =$ 0.08, consistent with the magnetic susceptibility data.  

We note that the differences in $T_{\text{co}}$ for the $x=0$ concentrations between magnetization and heat capacity data arise from crystals grown in separate batches.  These onset temperatures differ slightly from the formation of \textbf{q$_{\frac{1}{3}}$} order in our x-ray data (on a separate crystal) below 84 K, which is also reflective of a disorder-driven sample-to-sample variation.  All transitions reported here are below the nominal $T_{\text{co}}$=92 K reported in ScV$_6$Sn$_6$ \cite{Hasitha_2022}, suggesting small amounts of disorder introduced via the growth process can also perturb the stability of the charge ordered state.  
 
\begin{figure} 
	\centering
	\includegraphics[width=1\columnwidth]{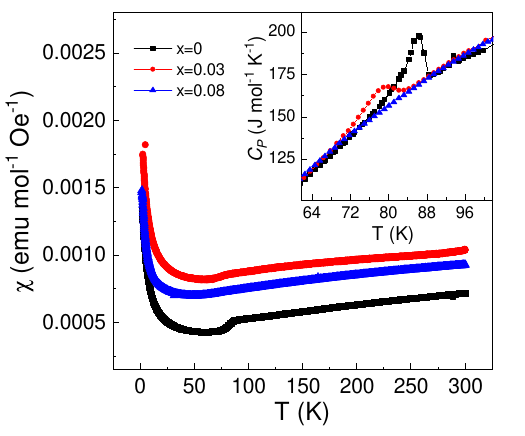}
	\caption{Temperature-dependent magnetic susceptibility, $\chi$ (T), of Sc$_{1-x}$Y$_x$V$_6$Sn$_6$ ($x=$ 0, 0.03, 0.08) single crystals measured under a magnetic field of $\mu_0 H$=1 T. The inset of shows heat capacity, C$_p$($T$), measurements of Sc$_{1-x}$Y$_x$V$_6$Sn$_6$ (x=0, 0.03, 0.08) single crystals. 
	}
	\label{}
\end{figure}

\begin{figure} 
	\centering
	\includegraphics[width=1\columnwidth]{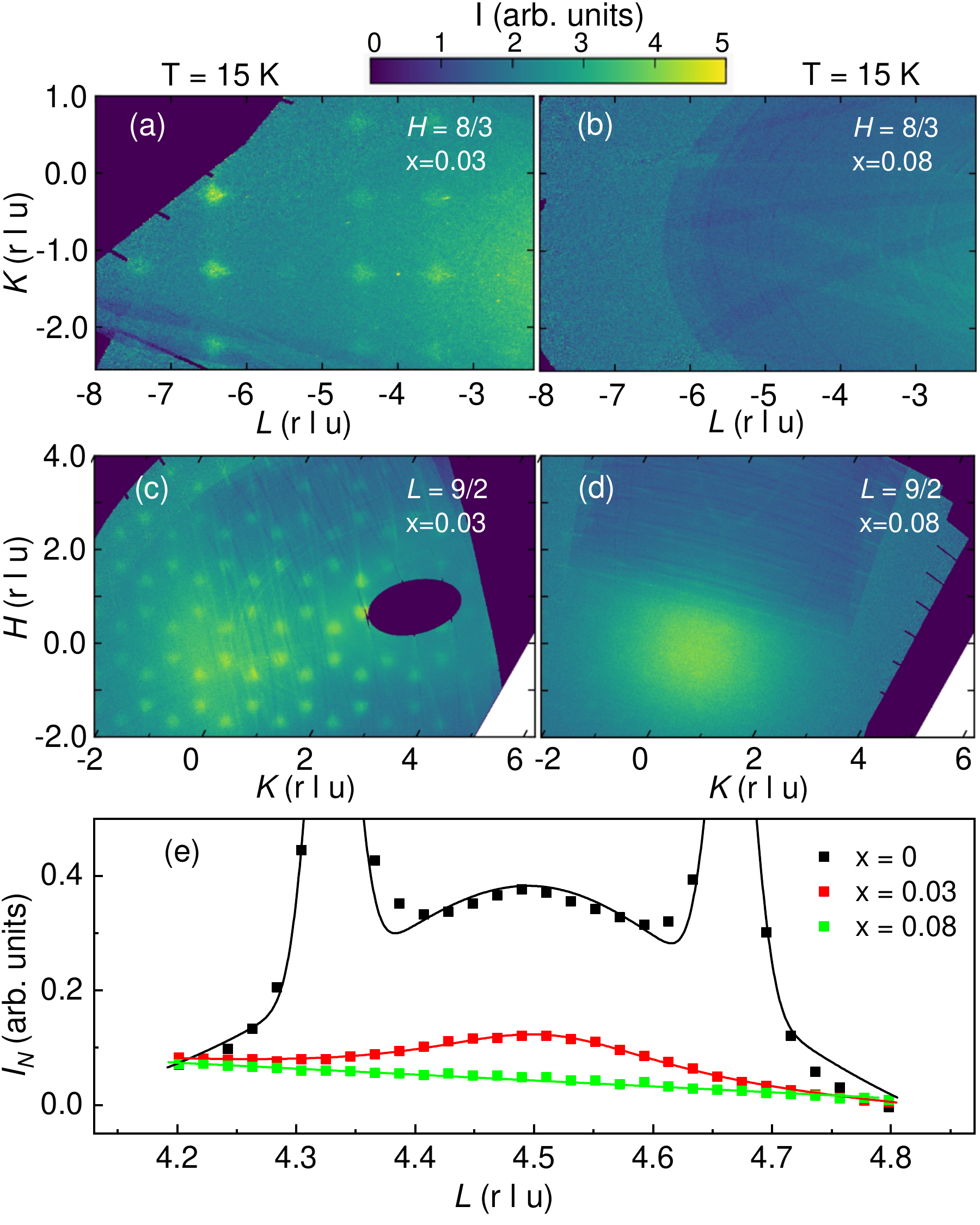}
	\caption{Synchrotron x-ray diffraction data collected on Sc$_{1-x}$Y$_x$V$_6$Sn$_6$ (x= 0.03, 0.08) crystals at 15 K.  Axis units are plotted in reciprocal lattice units (r l u). (a) and (c) show scattering in the ($\frac{8}{3}$, $K$, $L$) and ($H$, $K$, $\frac{9}{2}$) scattering planes, respectively, for the $x=$0.03 crystal at 15 K. (b) and (d) show scattering in the ($\frac{8}{3}$, $K$, $L$) and ($H$, $K$, $\frac{9}{2}$) scattering planes, respectively, for the $x=$0.08 crystal at 15 K. (e) $L$-cuts through the ($\frac{5}{3}$, $\frac{5}{3}$, $L$) position shows the evolution of \textbf{q$_{\frac{1}{2}}$} and \textbf{q$_{\frac{1}{3}}$} correlations in the $x=$ 0, 0.03, and 0.08 samples.
	}
	\label{Fig3}
\end{figure}

Fig \ref{Fig3} displays x-ray scattering data within the ($\frac{8}{3}$, $K$, $L$) and ($H$, $K$, $\frac{9}{2}$) planes of Sc$_{1-x}$Y$_x$V$_6$Sn$_6$ (x= 0.03, 0.08) for data collected at 15 K. For $x=0.03$, \textbf{q$_{\frac{1}{3}}$} order is largely suppressed and superlattice peaks are seemingly absent in the $L=\frac{1}{3}$ planes.  \textbf{q$_{\frac{1}{2}}$} correlations, however persist, consistent with notion of rapid, disorder-induced frustration of long-range \textbf{q$_{\frac{1}{3}}$} coherence, while the primary, frustrated \textbf{q$_{\frac{1}{2}}$} instability survives under light disorder.  Continued doping to $x=0.08$ reveals that all charge correlations vanish in both $L=\frac{1}{3}$ and $L=\frac{1}{2}$ planes.  The Y-doping evolution of both \textbf{q$_{\frac{1}{3}}$} and \textbf{q$_{\frac{1}{2}}$} type peaks normalized to a common Bragg peak are plotted in Figure 8 (c). 

\section{Discussion}

The persistence of \textbf{q$_{\frac{1}{2}}$} correlations to temperatures far above the onset of charge order and to higher impurity concentrations are consistent with it being the dominant lattice instability proposed in DFT calculations \cite{Binghai_2023}. This instability can be frustrated from forming correlations in-plane due to the repulsion of Sn1 sites from the V-plane, consistent with the fact that VSn is unstable in the FeSn structure and instead requires the added electrons from the interstitial $R$ sites to stabilize the derivative $R$V$_6$Sn$_6$ lattice.  

The smaller Sc promotes the Sn1-Sc-Sn1 trimer units to form along the $c$-axis-aligned chains through the hexagons in the kagome plane.  Replacement of a small percentage of Sc-sites with larger Y-sites (which do not promote trimer units as strongly) rapidly quenches charge order, consistent with a disrupted, one-dimensional Peierls-like instability along the chain.  The percolation limit of such a chain is generically expected to be small (or vanishing); for example, known examples in spin-Peierls chains reveal that nonmagnetic disorder at the level of $2\%$ is sufficient to disrupt long-range order \cite{PhysRevLett.80.4566}. Longer Y-Sn1 bonds that do not form movable trimers should amplify fluctuation effects along the chain at the expense of long-range order, consistent with our x-ray scattering results on Sc$_{1-x}$Y$_x$V$_6$Sn$_6$.  This also suggests that the remnant \textbf{q$_{\frac{1}{2}}$} fluctuations observed in the ground state of our $x=0$ sample derive from extrinsic disorder along the chains that also slightly suppresses the onset of \textbf{q$_{\frac{1}{3}}$} order.

The microscopic mechanism driving the formation of charge order is likely phonon softening along the Sc-Sn1-Sn1 chain-axis which drives the cooperative \textbf{q$_{\frac{1}{3}}$} distortion. The crucial role of phonons was demonstrated in recent time-resolved optics and ARPES data \cite{tuniz2023dynamics, hu2023phonon}.  Additional studies of ScV$_6$Sn$_6$ based on scanning tunneling microscopy (STM) and angle‐resolved photoemission spectroscopy (ARPES) \cite{cheng2023nanoscale,lee2023nature} also reveal a partial gap opening below the charge order transition; however our present results would suggest that some degree of partial gapping should remain above this transition due to the short-range order along \textbf{q$_{\frac{1}{2}}$} that persists to room temperature.  

While this paper was being finalized, a number of experiments exploring similar phenomena appeared.  Diffuse scattering was reported above the charge order transition in ScV$_6$Sn$_6$ in Saizheng Cao et al. \cite{cao2023competing} and A. Korshunov et al. \cite{korshunov2023softening} in agreement with the short-range correlations at \textbf{q$_{\frac{1}{2}}$} reported here.  Scattering at \textbf{q$_{\frac{1}{2}}$} in both of these studies, however, vanishes below $T_{\text{co}}$ which onsets at slightly higher temperatures that those reported here.  This is consistent with the notion of enhanced fluctuation effects arising from heightened disorder in the crystals studied in the present paper.  A recent transport study inferred the presence of a pseudogap above the charge order transition \cite{destefano2023pseudogap}, consistent with the presence of strong fluctuation effects above $T_{\text{co}}$ like those reported here.  Additionally, a recent preprint explored the impact of $R$-site disorder on the charge order transition and observed a rapid suppression \cite{meier2023tiny}, consistent with the results shown here. A microscopic model of the phase transitions in ScV$_6$Sn$_6$ has also been proposed \cite{hu2023kagome}.

\section{Conclusions} 

We have studied the structure and charge correlations in ScV$_6$Sn$_6$ at various temperatures and report an unusual interplay between $(\sqrt{3} \times \sqrt{3} \times 3)$ and $(\sqrt{3} \times \sqrt{3} \times 2)$  charge order instabilities.  Frustrated charge order at the \textbf{q$_{\frac{1}{2}}$} wave vector exhibits short-range correlations at high temperature that diverge upon cooling.  Once the inter-layer correlation length of \textbf{q$_{\frac{1}{2}}$} reaches three units cells, frustration is relieved via the onset of long-range \textbf{q$_{\frac{1}{3}}$} order.  Disorder introduced along the Sc-Sn1-Sn1 chains rapidly quenches both \textbf{q$_{\frac{1}{2}}$}- and \textbf{q$_{\frac{1}{3}}$}-type correlations, suggestive of a dominant one-dimensional chain instability driving charge order in this compound. A conceptual model describing the interplay between \textbf{q$_{\frac{1}{2}}$} and \textbf{q$_{\frac{1}{3}}$} correlations and the in-plane frustration of charge order was presented via consideration of the dominant bonding in the Sn1-Sc-Sn1 trimer units along the $c$-axis chains and an energy cost of Sn1 sites approaching the kagome planes.  Our results motivate future experiments under applied uniaxial pressure where the frustration of \textbf{q$_{\frac{1}{2}}$} charge order should promote an enhanced susceptibility to strain.

\begin{acknowledgments}
 S.D.W. acknowledges helpful discussions with Binghai Yan, Ram Seshadri, and Leon Balents. This work was supported by the National Science Foundation (NSF) through Enabling Quantum Leap: Convergent Accelerated Discovery Foundries for Quantum Materials Science, Engineering and Information (Q-AMASE-i): Quantum Foundry at UC Santa Barbara (DMR-1906325). The research made use of the shared facilities of the NSF Materials Research Science and Engineering Center at UC Santa Barbara (DMR- 1720256). The UC Santa Barbara MRSEC is a member of the Materials Research Facilities Network. (www.mrfn.org). Research conducted at the Center for High-Energy X-ray Science (CHEXS) is supported by the National Science Foundation (BIO, ENG and MPS Directorates) under award DMR-1829070.  K.M. and E.-A.K. are supported by the U.S.Department of Energy, Office of Basic Energy Sciences, Division of Materials Sciences and Engineering. The X-TEC analysis was carried out on a high-powered computing cluster funded in part by the New Frontier Grant from the College of Arts and Sciences at Cornell and by Gordon and Betty Moore Foundation’s EPiQS Initiative via grant GBMF10436. E.-A.K. was also supported by the Ewha Frontier 10-10 Research Grant. 
\end{acknowledgments}

%

\end{document}